
\input harvmac.tex
\Title{CTP/TAMU-34/92}{{Scattering of String Monopoles}
\footnote{$^\dagger$}{Work supported in part by NSF grant PHY-9106593.}}

\centerline{
Ramzi~R.~ Khuri\footnote{$^*$}{Supported by a World Laboratory Fellowship.}}
\bigskip\centerline{Center for Theoretical Physics}
\centerline{Texas A\&M University}\centerline{College Station, TX 77843}

\vskip .3in
In the low-velocity limit, multi-soliton solutions trace out geodesics
in the static solution manifold with distance defined by a metric on
moduli space. For the recently constructed multimonopole solutions of heterotic
string theory, we obtain a flat metric to leading order in the impact
parameter.
This result agrees with the trivial scattering predicted by a test monopole
calculation.

\Date{4/92}

\def\sqr#1#2{{\vbox{\hrule height.#2pt\hbox{\vrule width
.#2pt height#1pt \kern#1pt\vrule width.#2pt}\hrule height.#2pt}}}

\def\met {g_{\mu\nu}}

\def\b#1{\vec\beta_{#1}}

\lref\reyone {S.--J.~Rey {\it Axionic String Instantons
and their Low-Energy Implications}, Proceedings, Tuscaloosa 1989,
Superstrings and particle theory, p.291.}

\lref\reytwo {S.--J.~Rey, Phys. Rev. {\bf D43} (1991) 526.}

\lref\abenone{I.~Antoniadis, C.~Bachas, J.~Ellis and D.~V.~Nanopoulos,
Phys. Lett. {\bf B211} (1988) 393.}

\lref\abentwo{I.~Antoniadis, C.~Bachas, J.~Ellis and D.~V.~Nanopoulos,
Nucl. Phys. {\bf B328} (1989) 117.}

\lref\mtone{R.~R.~Metsaev and A.~A.~Tseytlin, Phys. Lett.
{\bf B191} (1987) 354.}

\lref\mttwo{R.~R.~Metsaev and A.~A.~Tseytlin,
Nucl. Phys. {\bf B293} (1987) 385.}

\lref\cfmp{C.~G.~Callan, D.~Friedan, E.~J.~Martinec
and M.~J.~Perry, Nucl. Phys. {\bf B262} (1985) 593.}

\lref\ckp{C.~G.~Callan,
I.~R.~Klebanov and M.~J.~Perry, Nucl. Phys. {\bf B278} (1986) 78.}

\lref\love{C.~Lovelace, Phys. Lett. {\bf B135} (1984) 75.}

\lref\fridven{B.~E.~Fridling and A.~E.~M.~Van de Ven,
Nucl. Phys. {\bf B268} (1986) 719.}

\lref\gepwit{D.~Gepner and E.~Witten, Nucl. Phys. {\bf B278} (1986) 493.}

\lref\quartet{D.~J.~Gross,
J.~A.~Harvey, E.~J.~Martinec and R.~Rohm, Nucl. Phys. {\bf B267} (1986) 75.}

\lref\dine{M.~Dine, Lectures delivered at
TASI 1988, Brown University (1988) 653.}

\lref\brone{E.~A.~Bergshoeff and M.~de Roo, Nucl.
Phys. {\bf B328} (1989) 439.}

\lref\brtwo{E.~A.~Bergshoeff and M.~de Roo, Phys. Lett. {\bf B218} (1989) 210.}

\lref\chsone{C.~G.~Callan, J.~A.~Harvey and A.~Strominger, Nucl. Phys.
{\bf B359} (1991) 611.}

\lref\chstwo{C.~G.~Callan, J.~A.~Harvey and A.~Strominger, Nucl. Phys.
{\bf B367} (1991) 60.}

\lref\bpst{A.~A.~Belavin, A.~M.~Polyakov, A.~S.~Schwartz and Yu.~S.~Tyupkin,
Phys. Lett. {\bf B59} (1975) 85.}

\lref\thooft{G.~'t~Hooft, Nucl. Phys. {\bf B79} (1974) 276.}

\lref\hoofan{G.~'t~Hooft, Phys. Rev. Lett., {\bf 37} (1976) 8.}

\lref\wil{F.~Wilczek, in {\it Quark confinement and field theory},
Ed. D.~Stump and D.~Weingarten, John Wiley and Sons, New York (1977).}

\lref\cofa{E.~Corrigan and D.~B.~Fairlie, Phys. Lett. {\bf B67} (1977) 69.}

\lref\jackone{R.~Jackiw, C.~Nohl and C.~Rebbi, Phys. Rev. {\bf D15} (1977)
1642.}

\lref\jacktwo{R.~Jackiw, C.~Nohl and C.~Rebbi, in {\it Particles and
Fields}, Ed. David Boal and A.~N.~Kamal, Plenum Publishing Co., New York
(1978), p.199.}

\lref\rkinst{R.~R.~Khuri, Phys. Lett. {\bf B259} (1991) 261.}

\lref\rkscat{C.~G.~Callan and R.~R.~Khuri, Phys. Lett. {\bf B261} (1991) 363.}

\lref\rkmant{R.~R.~Khuri, {\it Manton Scattering of String Solitons}
PUPT-1270 (to appear in Nucl. Phys. {\bf B}).}

\lref\rkdg{R.~R.~Khuri, {\it Some Instanton Solutions in String
Theory} to appear in Proceedings of the XXth International Conference on
Differential Geometric Methods in Theoretical Physics, World Scientific,
October 1991.}

\lref\rkthes{R.~R.~Khuri, {\it Solitons and Instantons in String Theory},
 Princeton University Doctoral Thesis, August 1991.}

\lref\rksing{M.~J.~Duff, R.~R.~Khuri and J.~X.~Lu, {\it String and
Fivebrane Solitons: Singular or Non-singular?}, Texas A\&M preprint,
CTP/TAMU-89/91 (to appear in Nucl. Phys. {\bf B}).}

\lref\rkorb{R.~R.~Khuri and H.~S.~La, {\it Orbits of a String around a
Fivebrane}, Texas A\&M preprint, CTP/TAMU-95/91
(submitted to Phys. Rev. Lett.).}

\lref\rkmot{R.~R.~Khuri and H.~S.~La, {\it String Motion in Fivebrane
Geometry}, Texas A\&M preprint, CTP/TAMU-98/91 (submitted to Nucl. Phys. B).}

\lref\rkmonex{R.~R.~Khuri {\it A Heterotic Multimonopole Solution},
Texas A\&M preprint, CTP/TAMU-35/92.}

\lref\rkmono{R.~R.~Khuri {\it A Multimonopole Solution in
String Theory}, Texas A\&M preprint, CTP/TAMU-33/92.}

\lref\rkmscat{R.~R.~Khuri {\it Scattering of String Monopoles},
Texas A\&M preprint, CTP/TAMU-34/92.}

\lref\ginsp{P.~Ginsparg, Lectures delivered at
Les Houches summer session, June 28--August 5, 1988.}

\lref\swzw {W.~Boucher, D.~Friedan and A.~Kent, Phys. Lett.
{\bf B172} (1986) 316.}

\lref\dghrr{A.~Dabholkar, G.~Gibbons, J.~A.~Harvey and F.~Ruiz Ruiz,
Nucl. Phys. {\bf B340} (1990) 33.}

\lref\dabhar{A.~Dabholkar and J.~A.~Harvey,
Phys. Rev. Lett. {\bf 63} (1989) 478.}

\lref\prso{M.~K.~Prasad and C.~M.~Sommerfield, Phys. Rev. Lett. {\bf 35}
(1975) 760.}

\lref\jim{J.~A.~Harvey and J.~Liu, Phys. Lett. {\bf B268} (1991) 40.}

\lref\mantone{N.~S.~Manton, Nucl. Phys. {\bf B126} (1977) 525.}

\lref\manttwo{N.~S.~Manton, Phys. Lett. {\bf B110} (1982) 54.}

\lref\mantthree{N.~S.~Manton, Phys. Lett. {\bf B154} (1985) 397.}

\lref\atiyah{M.~F.~Atiyah and N.~J.~Hitchin, Phys. Lett. {\bf A107}
(1985) 21.}

\lref\atiyahbook{M.~F.~Atiyah and N.~J.~Hitchin, {\it The Geometry and
Dynamics of Magnetic Monopoles}, Princeton University Press, 1988.}

\lref\strom{A.~Strominger, Nucl. Phys. {\bf B343} (1990) 167.}

\lref\gsw{M.~B.~Green, J.~H.~Schwartz and E.~Witten,
{\it Superstring Theory} vol. 1, Cambridge University Press (1987).}

\lref\polch{J.~Polchinski, Phys. Lett. {\bf B209} (1988) 252.}

\lref\dfluone{M.~J.~Duff and J.~X.~Lu, Nucl. Phys. {\bf B354} (1991) 141.}

\lref\dflutwo{M.~J.~Duff and J.~X.~Lu, Nucl. Phys. {\bf B354} (1991) 129.}

\lref\dfluthree{M.~J.~Duff and J.~X.~Lu, Phys. Rev. Lett. {\bf 66}
(1991) 1402.}

\lref\dflufour{M.~J.~Duff and J.~X.~Lu, Nucl. Phys. {\bf B357} (1991)
534.}

\lref\dfstel{M.~J.~Duff and K.~S.~Stelle, Phys. Lett. {\bf B253} (1991)
113.}

\lref\ferone{R.~C.~Ferrell and D.~M.~Eardley, Phys. Rev. Lett. {\bf 59}
(1987) 1617.}

\lref\fertwo{R.~C.~Ferrell and D.~M.~Eardley, {\it Slowly Moving
Maximally Charged Black Holes} in Frontiers in Numerical Relativity,
Cambridge University Press, 1987.}

\lref\gh{G.~W.~Gibbons and S.~W.~Hawking, Phys. Rev. {\bf D15}
(1977) 2752.}

\lref\ghp{G.~W.~Gibbons, S.~W.~Hawking and M.~J.~Perry, Nucl. Phys. {\bf B318}
(1978) 141.}

\lref\briho{D.~Brill and G.~T.~Horowitz, Phys. Lett. {\bf B262} (1991)
437.}

\lref\gidone{S.~B.~Giddings and A.~Strominger, Nucl. Phys. {\bf B306}
(1988) 890.}

\lref\gidtwo{S.~B.~Giddings and A.~Strominger, Phys. Lett. {\bf B230}
(1989) 46.}

\lref\raj{R.~Rajaraman, {\it Solitons and Instantons}, North Holland,
1982.}

\lref\chsw{P.~Candelas, G.~T.~Horowitz, A.~Strominger and E.~Witten,
Nucl. Phys. {\bf B258} (1984) 46.}

\lref\bogo{E.~B.~Bogomolnyi, Sov. J. Nucl. Phys. {\bf 24} (1976) 449.}

\lref\cogo{E.~Corrigan and P.~Goddard, Comm. Math. Phys. {\bf 80} (1981)
575.}

\lref\wardone{R.~S.~Ward, Comm. Math. Phys. {\bf 79} (1981) 317.}

\lref\wardtwo{R.~S.~Ward, Comm. Math. Phys. {\bf 80} (1981) 563.}

\lref\wardthree{R.~S.~Ward, Phys. Lett. {\bf B158} (1985) 424.}

\lref\groper{D.~J.~Gross and M.~J.~Perry, Nucl. Phys. {\bf B226} (1983)
29.}

\lref\ash{{\it New Perspectives in Canonical Gravity}, ed. A.~Ashtekar,
Bibliopolis, 1988.}

\lref\lich{A.~Lichnerowicz, {\it Th\' eories Relativistes de la
Gravitation et de l'Electro-magnetisme}, (Masson, Paris 1955).}

\lref\goldstein{H.~Goldstein, {\it Classical Mechanics}, Addison-Wesley,
1981.}

\lref\dflufive{M.~J.~Duff and J.~X.~Lu, Class. Quant. Grav. {\bf 9}
(1992) 1.}

\lref\dflusix{M.~J.~Duff and J.~X.~Lu, Phys. Lett. {\bf B273} (1991)
409.}

\lref\hlp{J.~Hughes, J.~Liu and J.~Polchinski, Phys. Lett. {\bf B180}
(1986).}

\lref\town{P.~K.~Townsend, Phys. Lett. {\bf B202} (1988) 53.}

\lref\duff{M.~J.~Duff, Class. Quant. Grav. {\bf 5} (1988).}

\lref\rossi{P.~Rossi, Physics Reports, 86(6) 317-362.}

\lref\gksone{B.~Grossman, T.~W.~Kephart and J.~D.~Stasheff, Commun. Math.
Phys. {\bf 96} (1984) 431.}

\lref\gkstwo{B.~Grossman, T.~W.~Kephart and J.~D.~Stasheff, Commun. Math.
Phys. {\bf 100} (1985) 311.}

\lref\gksthree{B.~Grossman, T.~W.~Kephart and J.~D.~Stasheff, Phys. Lett.
{\bf B220} (1989) 431.}

\newsec{Introduction}

In recent work\refs{\rkmono} an exact multimonopole solution of
heterotic string theory was presented. This solution was obtained via a
modification of the 't Hooft ansatz\refs{\hoofan\wil\cofa\jackone{--}\jacktwo}
for the Yang-Mills instanton. An analogous solution in Yang-Mills field
theory saturates a Bogomoln'yi\bogo\ bound and possesses the topology and far
field limit of a multimonopole configuration, but has divergent action near
each source. In the string solution, however, the divergences from the
Yang-Mills sector are precisely cancelled by those from the gravity sector,
so that the action is finite and easily computed\rkmono. In this
letter, we study the dynamics of the string monopoles and find that, unlike
BPS\refs{\prso,\bogo}\ monopoles, the string monopoles scatter trivially to
leading order in the impact parameter.

We study the scattering of two string monopoles by two methods. The first
approach computes the Manton metric on moduli space, which defines distance on
the static solution manifold. We first invert the $O(\beta)$ time-dependent
constraint equations and replace the solution into the action. The resultant
kinetic action defines the metric on moduli space. A flat metric is obtained
to leading order in the impact parameter, a result which implies trivial
scattering between string monopoles.

An independent calculation of the dynamic force on a test string monopole
moving in the background of a source string monopole yields a zero dynamic
force to lowest order in the velocity, again implying trivial scattering.
This computation thus confirms the flat metric result.

\newsec{Manton Metric for String Monopoles}

The bosonic fields for the exact self-dual multimonopole solution of heterotic
string theory with zero background fermi fields are given by\rkmono
\eqn\anstz{\eqalign{\met&=e^{2\phi}\delta_{\mu\nu},\quad g_{ab}=\eta_{ab},\cr
H_{\mu\nu\lambda}&=\pm\epsilon_{\mu\nu\lambda\sigma}\partial^\sigma\phi,\cr
e^{2\phi}&=e^{2\phi_0}f,\cr
A_\mu&=i \overline{\Sigma}_{\mu\nu}\partial_\nu \ln f,\cr}}
where $\mu,\nu,\lambda,\sigma=1,2,3,4$, $a,b=0,5,6,7,8,9$,
$\overline{\Sigma}_{\mu\nu}=\overline{\eta}^{i\mu\nu}(\sigma^i/2)$
for $i=1,2,3$ ($\sigma^i$, $i=1,2,3$ are the $2\times 2$ Pauli matrices) where
\eqn\hfeta{\eqalign{\overline{\eta}^{i\mu\nu}=-\overline{\eta}^{i\nu\mu}
&=\epsilon^{i\mu\nu},\qquad\qquad \mu,\nu=1,2,3,\cr
&=-\delta^{i\mu},\qquad\qquad \nu=4 \cr}}
and where
\eqn\fdmono{f=1+\sum_{n=1}^N{m_n\over |\vec x - \vec a_n|},}
where $m_n$ is the charge and $\vec a_n$ the location in
the three-space $(123)$ of the $n$th monopole. The anti-self-dual solution is
similar, with the $\delta$-term in \hfeta\ changing sign. This solution
was shown to have multimonopole structure\rkmono\ in the three-space,
each source having topological charge $Q=1$ and magnetic charge $m=1/g$,
where $g$ is the YM coupling constant.

If we make the identification $\Phi\equiv A_4$, then the gauge and Higgs
fields may be simply written in terms of the dilaton as
\eqn\stmono{\eqalign{\Phi^a&=-{2\over g}\delta^{ia}\partial_i\phi,\cr
A_k^a&=-{2\over g}\epsilon^{akj}\partial_j\phi\cr}}
for the self-dual solution. For the anti-self-dual solution, the Higgs
field simply changes sign. A toroidal compactification along the lines
of \jim\ can be adopted, so that we consider the dynamics of our solution
in the four-dimensional spacetime $(0123)$. As usual, the existence of a
static multi-soliton solution depends on the ``zero force'' condition.

Owing to the exactness condition
$A_\mu=\Omega_{\pm\mu}$\refs{\dine,\brone,\brtwo}\ (where
$\Omega_{\pm\mu}$ is the generalized connection defined in
\rkmono), the higher order in $\alpha'$ terms drop out from the action,
and the static multimonopole mass can be computed from the tree-level
action\refs{\rkmant,\rkmono}
\eqn\trueaction{S=-{1\over 2\kappa^2}\left[\int dt\left(\int d^3x
\sqrt{g} e^{-2\phi}\left( R + 4(\nabla\phi)^2 - {H^2\over 12}\right)
+2\int_{\partial M}\left(e^{-2\phi}K-K_0\right)\right)\right],}
where we have added a Gibbons-Hawking surface term (GHST) to cancel the
double derivative terms in the action\refs{\gh,\ghp,\gidone,\briho,\rkmant}.
$\partial M$ is the surface boundary and $K$ and $K_0$ are the
traces of the fundamental form of the boundary surface embedded in the
metric $g$ and the Minkowskian metric $\eta$ respectively. The addition of a
surface term does not, of course, affect the equations of motion.
The multimonopole mass is given by\rkmono
\eqn\totmass{M_T={8\pi\over \kappa^2}\sum_{n=1}^N m_n,}
where $m_n=1/g$ for $n=1,2...N$.

We wish to study dynamics of the string monopoles.
Manton's prescription\manttwo\ for the study of soliton scattering may be
summarized as follows. We first invert the constraint equations of the system.
The resultant time dependent field configuration does not in general
satisfy the full time dependent field equations, but provides
an initial data point for the fields and their time derivatives.
Another way of saying this is that the initial motion is tangent to the
set of exact static solutions.  The kinetic action obtained
by replacing the solution to the constraints into the action defines a
metric on the parameter space of static solutions. This metric defines
geodesic motion on the moduli space\manttwo.

A calculation of the metric on moduli space for the scattering of BPS
monopoles and a description of its geodesics was worked out by Atiyah
and Hitchin\atiyah. Several interesting properties of monopole
scattering were found, such as the conversion of monopoles into dyons
and the right angle scattering of two monopoles on a direct collision
course\refs{\atiyah,\atiyahbook}. The configuration space is found to
be a four-dimensional manifold $M_2$ with a self-dual Einstein metric.

In this section, we adapt Manton's prescription to study the
dynamics of heterotic string monopoles. A similar procedure was
followed in \rkmant\ for the Manton scattering of heterotic instantons.
Indeed, many of the formal computations carry over from the instanton
computation. For the monopoles, however, the divergences plagueing the
instanton calculation are absent, thus rendering our task far simpler.
In both cases, we follow essentially
the same steps that Manton outlined for monopole scattering, but take
into account the peculiar nature of the string effective action. Since
we work in the low-velocity limit, our kinematic analysis is nonrelativistic.

We first solve the constraint equations for the soliton solutions.
These equations are simply the $(0j)$
components of the equations of motion (see \refs{\rkinst,\rkmant})
\eqn\constraints{\eqalign{R_{0j}-{1\over
4}H^2_{0j}+2\nabla_0\nabla_j\phi&=0,\cr
-{1\over 2}\nabla_kH^k{}_{0j}+H_{0j}{}^k\partial_k\phi&=0.\cr}}
Note that we use the tree-level equations of motion, as the higher order
corrections in $\alpha'$ automatically vanish.
We wish to find an $O(\beta)$ solution to the above equations which
represents a quasi-static version of \anstz\ (i.e. a solution of
the form \anstz\ but with time dependent $\vec a_i$). In other words,
we would like to give each source an arbitrary transverse velocity
$\vec\beta_n$ in the $(123)$ subspace of the four-dimensional transverse space
and see what corrections to the fields are required by the
constraints. The vector $\vec a_n$ representing the position of source
$n$ in the three-space $(123)$ is given by
\eqn\aunty{\vec a_n(t)=\vec A_n + \vec\beta_nt,}
where $\vec A_n$ is the initial position of the $n$th source. Note that at
$t=0$ we have an exact static multi-soliton solution. Our solution to
the constraints will adjust our quasi-static approximation so that the
initial motion in the parameter space is tangent to the initial
exact solution at $t=0$.

The $O(\beta)$ solution to the constraints is given by
\eqn\orderbeta{\eqalign{e^{2\phi(\vec x,t)}&=1+\sum_{n=1}^N{m_n\over
|\vec x - \vec a_n(t)|},\cr g_{00}&=-1,\qquad g^{00}=-1,\qquad
g_{ij}=e^{2\phi}\delta_{ij},\qquad g^{ij}=e^{-2\phi}\delta_{ij},\cr
g_{0i}&=-\sum_{n=1}^N{m_n\vec\beta_n\cdot \hat x_i\over |\vec x - \vec
a_n(t)|},\qquad g^{0i}=e^{-2\phi}g_{0i},\cr
H_{ijk}&=\epsilon_{ijkm}\partial_m e^{2\phi},\cr
H_{0ij}&=\epsilon_{ijkm}\partial_m g_{0k}=\epsilon_{ijkm}\partial_k
\sum_{n=1}^N{m_n\vec\beta_n\cdot \hat x_m\over |\vec x - \vec a_n(t)|},\cr}}
where $i,j,k,m=1,2,3,4$, the $\vec a_n(t)$ are given by \aunty\ and we use a
flat space $\epsilon$-tensor. Note that $g_{00}$, $g_{ij}$ and $H_{ijk}$ are
unaffected to order $\beta$. Also note that we can interpret the solitons
as either line sources in the four-dimensional space $(1234)$ or point
sources in the three-dimensional subspace $(123)$.

The kinetic Lagrangian is obtained by replacing the expressions for the
fields in \orderbeta\ into \trueaction. Since \orderbeta\ is a solution to
order $\beta$, the leading order terms in the action (after the
quasi-static part) are of order $\beta^2$. In the volume term of the
action, $O(\beta)$ terms in the solution give $O(\beta^2)$ terms in the
kinetic action. As explained in \rkmant, the contribution of the GHST
to the kinetic action can be written in the form $m_s{\beta^2}/2$ for
each source, and the contributions of the sources can be simply
added. The GHST does not therefore play an important role in the dynamics
of the string monopoles, but merely serves to give the correct total mass.
Collecting all $O(\beta^2)$ terms in $S_V$ we
get the following kinetic Lagrangian density for the volume term:
\eqn\kinlag{\eqalign{{\cal L}_{kin}=-{1\over 2\kappa^2}\Biggl(
&4\dot \phi\vec M\cdot\vec \nabla\phi
-e^{-2\phi}\partial_iM_j\partial_iM_j
-e^{-2\phi}M_k\partial_j\phi\left(\partial_jM_k-\partial_kM_j\right)\cr
&+4M^2e^{-2\phi}(\vec \nabla\phi)^2
+2\partial_t^2e^{2\phi}-4\partial_t(\vec M\cdot\vec \nabla\phi)-4\vec
\nabla\cdot(\dot\phi\vec M)\Biggr),\cr}}
where $\vec M\equiv -\sum_{n=1}^N{m_n\vec \beta_n\over |\vec x - \vec
a_n(t)|}$. Henceforth let $\vec X_n\equiv \vec x - \vec a_n(t)$.
The last three terms in \kinlag\ are time-surface or space-surface terms
which vanish when integrated. Note that the kinetic Lagrangian has the
same form as in \rkmant. The contributions of the GHST are again simply
flat kinetic terms.

In contrast to the instanton case, the kinetic Lagrangian
$L_{kin}=\int d^3x{\cal L}_{kin}$ for monopole scattering converges
everywhere. This can be seen simply by studying the limiting behaviour
of $L_{kin}$ near each source. For a single source at $r=0$ with magnetic
charge $m$ and velocity $\beta$, we collect the logarithmically divergent
pieces
and find that they cancel:
\eqn\logdiv{{m\beta^2\over 2}\int r^2 drd\theta \sin\theta d\phi
\left(-{1\over r^3} + {3\cos^2\theta\over r^3}\right)=0.}
So unlike the instanton case, in which we were compelled to extract
information from the convergent interaction terms, in this case we can
use the self-terms directly.

We now specialize to the case of two heterotic monopoles of magnetic
charge $m_1=m_2=m=1/g$ and velocities $\vec\beta_1$ and $\vec\beta_2$.
Let the monopoles be located at $\vec a_1$ and $\vec a_2$.
Our moduli space consists of the configuration space of the relative
separation vector $\vec a\equiv \vec a_2 - \vec a_1$.
The most general kinetic Lagrangian can be written as
\eqn\genkinlag{\eqalign{L_{kin}=&h(a)(\b1\cdot\b1+\b2\cdot\b2)+p(a)\left(
(\b1\cdot\hat a)^2 + (\b2\cdot\hat a)^2\right)\cr
&+2f(a)\b1\cdot\b2 + 2g(a)(\b1\cdot\hat a)(\b2\cdot\hat a).\cr}}
Now suppose $\b1 = \b2 =\vec\beta$, so that \genkinlag\ reduces to
\eqn\boostlag{L_{kin}=(2h+2f)\beta^2+(2p+2g)(\vec\beta\cdot\hat a)^2.}
This configuration, however, represents the boosted solution of the
two-static soliton solution. The kinetic energy should therefore be
simply
\eqn\cmke{L_{kin}={M_T\over 2}\beta^2,}
where $M_T=M_1+M_2=2M={16\pi m}/{\kappa^2}$ is the total mass of
the two soliton solution. It then follows that the anisotropic part of
\boostlag\ vanishes and we have
\eqn\hfpg{\eqalign{g+p&=0,\cr 2(h+f)&={M_T\over 2}.\cr}}

It is therefore sufficient to compute $h$ and $p$. This can be done by
setting $\vec\beta_1=\vec\beta$ and $\vec\beta_2=0$.
The kinetic Lagrangian then reduces to
\eqn\rdkinlag{L_{kin}=h(a)\beta^2 + p(a)(\vec\beta\cdot\hat a)^2.}
Suppose for simplicity
also that $\vec a_1=0$ and $\vec a_2=\vec a$ at $t=0$.
The Lagrangian density of the volume term in this case is given by
\eqn\voltm{\eqalign{{\cal L}_{kin}&={-1\over 2\kappa^2}\Biggl(
{3m^3 e^{-4\phi}\over 2r^4}(\vec\beta\cdot\vec x)\left[
{\vec\beta\cdot\vec x\over r^3} + {\vec\beta\cdot(\vec x-\vec a)
\over |\vec x-\vec a|^3}\right] - {e^{-2\phi}m^2\beta^2\over r^4}\cr
&-{e^{-4\phi}m^3\beta^2\over 2r^4}\left( {1\over r} +
{\vec x\cdot(\vec x-\vec a)\over |\vec x-\vec a|^3}\right) +
{e^{-6\phi}m^4\beta^2\over r^2}\left( {1\over r^4} + {1\over |\vec x-\vec a|^4}
+ {2\vec x\cdot(\vec x-\vec a)\over r^3|\vec x-\vec a|^3}\right)\Biggr).\cr}}
The GHST contribution to the kinetic Lagrangian can be simply added
after integration and will not affect the analysis below.

The integration of the kinetic Lagrangian density in \voltm\ over three-space
yields the kinetic Lagrangian from which the metric on moduli space can be
read off. For large $a$, the nontrivial leading order  behaviour of the
components of the metric, and hence for the functions $h(a)$ and $p(a)$, is
generically of order $1/a$. In fact, for Manton scattering of YM monopoles,
the leading order scattering angle is $2/b$\mantthree, where $b$ is the impact
parameter. In this paper, we restrict our computation to the leading order
metric in moduli space. A tedious but straightforward collection of $1/a$
terms in the Lagrangian yields
\eqn\leadi{{-1\over 2\kappa^2}{1\over a}\int d^3x\left[ -{3m^4e^{-6\phi_1}
\over r^7}(\vec\beta\cdot\vec x)^2 + {m^3e^{-4\phi_1}\over r^4}\beta^2+
{m^4e^{-6\phi_1}\over r^5}\beta^2 - {3m^5e^{-8\phi_1}\over r^6}\beta^2
\right],}
where $e^{2\phi_1}\equiv 1+m/r$.
The first and third terms clearly cancel after integration over three-space.
The second and fourth terms are spherically symmetric. A simple integration
yields
\eqn\leadii{\int_0^\infty r^2dr \left( {e^{-4\phi_1}\over r^4} -
{3m^2e^{-8\phi_1}\over r^6}\right)
=\int_0^\infty {dr\over (r+m)^2} - 3m^2\int_0^\infty {dr\over (r+m)^4}=0.}
The $1/a$ terms therefore cancel, and the leading order metric on moduli
space is flat. This implies that the leading order scattering is trivial.
In other words, there is no deviation from the initial trajectories to
leading order in the impact parameter.

The above result is rather surprising and suggests that, in addition
to the static force, the leading order dynamic force also vanishes.
For pure YM monopoles, this is certainly not the case. For the string
monopoles, however, the dynamic YM force is precisely cancelled by the
dynamic gravity sector force. In the next section, we adopt a different
approach to the computation of the dynamic force in order to confirm the
flat metric result.

\newsec{Test Monopole Calculation}

We now employ the test-soliton approach of
\refs{\dghrr,\rkscat} to compute the dynamic force exerted on a
test string monopole moving in the background of a source string monopole.
Again only the massless fields in the gravitational sector come in to
play at tree-level. Since the monopoles have fivebrane structure, we
adopt the fivebrane action of Duff and Lu\refs{\dfluone,\dflutwo}
\eqn\sigfiv{\eqalign{S_{\sigma_5}=&-T_6\int d^6\xi\Biggl({1\over 2}
\sqrt{-\gamma}\gamma^{mn}\partial_mX^M\partial_nX^N g_{MN}e^{-\phi/6}
-2\sqrt{-\gamma}\cr&+{1\over 6!}\epsilon^{mnpqrs}\partial_mX^M\partial_nX^N
\partial_pX^P\partial_qX^Q\partial_rX^R\partial_sX^S A_{MNPQRS}\Biggr),\cr}}
where $m,n,p,q,r,s=0,5,6,7,8,9$ are fivebrane indices and
$M,N,P,Q,R,S=0,1,...9$ are spacetime indices
(transverse indices are denoted by $i,j=1,2,3,4$).
$\gamma_{mn}$ is a $5+1$-dimensional worldsheet metric, $g_{MN}$ is the
canonical spacetime metric and $A_{MNPQRS}$ is the
antisymmetric six-form potential whose curl $K=dA$ is
dual to the antisymmetric field strength $H_{\alpha\beta\gamma}$.

The multimonopole solution written in this frame is given by
\eqn\fivanstz{\eqalign{ds^2&=e^{2A}\eta_{mn}dx^mdx^n+e^{2B}\delta_{ij}
dx^idx^j, \cr  A_{056789}&=-e^C,\cr}}
where all other components of $A_{MNPQRS}$ are set to zero and the dilaton
$\phi$ and the scalar functions $A$, $B$ and $C$ are given by
\eqn\abc{\eqalign{A&=-{(\phi-\phi_0)\over 4},\cr
B&={3(\phi-\phi_0)\over 4},\cr C&=-2\phi+{3\phi_0\over 2},\cr}}
where $\phi_0$ is the value of the dilaton field at infinity and
\eqn\fivsltn{e^{2\phi}=e^{2\phi_0}\left(1+\sum_{n=1}^N
{m_n\over |\vec x -\vec a_n|}\right),}
where $\vec x$ and $\vec a_n$ are again vectors in the three-dimensional
subspace $(123)$ of the transverse space $(1234)$.

The Lagrangian for a test monopole moving in a background of identical static
source monopoles is given by substituting
\fivanstz\ in \sigfiv\ and then eliminating the worldbrane metric. The result
is
\eqn\fbworldsheet{{\cal L}_6=-T_6\left[\sqrt{-\det (e^{-2\phi/3+\phi_0/2}
\eta_{mn}+e^{4\phi/3-3\phi_0/2}\partial_m X^M\partial_n X_M)}
-e^{-2\phi+3\phi_0/2}\right].}

Since the test-monopole moves only in the $(123)$ subspace of the
transverse space (there is no motion along or field dependence on the
direction $x_4$), \fbworldsheet\ reduces in the low-velocity limit to
\eqn\fblagrange{\eqalign{{\cal L}_6&\simeq-T_6\left[e^{-2\phi+3\phi_0/2}
\left(1-\half e^{2(\phi-\phi_0)}(\dot X^i)^2\right)
-e^{-2\phi+3\phi_0/2}\right]\cr
&={T_6\over 2}e^{-\phi_0/2}(\dot X^i)^2~,\cr}}
where $i=1,2,3$.
Again both the static force and the nontrivial $O(v^2)$ velocity-dependent
force vanish. Hence this result also predicts trivial scattering, in direct
agreement with the flat Manton metric calculation.

\newsec{Conclusion}

In \rkmono, an exact multimonopole solution of heterotic string theory
was presented. An analogous solution in YM field theory was found to
have divergent action near each source. In the string theory solution,
however, the divergences from the Yang-Mills sector are exactly cancelled by
divergences in the gravity sector. The cancellation between the gauge and
gravitational sectors also influences the dynamics of the string monopoles.
In this paper, we found from both a Manton metric on moduli space calculation
and a test string monopole calculation that the leading order dynamic force
between two string monopoles vanishes. This result implies trivial scattering
between string monopoles to leading order in the impact parameter in the
low-velocity limit.

\vfil\eject
\listrefs
\bye